\documentstyle[A4,12pt]{article}

\newcommand{\spao}[1]{\mbox{\hspace{#1}}}
\newcommand{\spav}[1]{\parbox{1mm}{\vspace*{#1}}}
\newcommand{\om}{\Omega}

\newcommand{\os}{\Omega^{\ast}}
\newcommand{\oms}{\Omega^{\ast}}

\newcommand{\drp}{\partial}
\newcommand{\drpb}{\bar{\partial}}

\newcommand{\zb}{\bar{z}}

\newcommand{\bc}{\mu^z_{\bar{z}}}
\newcommand{\bbc}{\mu^{zz}_{\bar{z}}}
\newcommand{\ssc}{\rho_{zz}}
\newcommand{\rz}{\rho_{zzz}}
\newcommand{\Lb}{\Lambda_{ab}}
\newcommand{\cz}{c^z}
\newcommand{\czz}{c^{zz}}
\newcommand{\dd}{\frac{1}{2}}
\newcommand{\dt}{\frac{1}{3}}
\newcommand{\f}[1]{\frac{#1}{3}}
\newcommand{\ds}{\frac{1}{6}}
\newcommand{\ca}{{\cal A}}
\newcommand{\cd}{\bar{{\cal D}}}

\begin{document}





\hskip 10cm {\footnotesize  DSF-T-95/12}

\hskip 10cm {\footnotesize  INFN NA-IV-95/12}

\hskip 10cm {\footnotesize  CPTMB/PT/95-2/}
\vskip 2cm
\begin{center}
\spav{5mm}\\
{\LARGE\bf Consistent anomalies of the induced W gravities
$^{\dagger}$\\ }
\spav{1.cm}\\
{\large Mario Abud $^{(1)}$, Jean-Pierre Ader $^{(2)}$ and Luigi
Cappiello$^{(1)}$}\\
\spao{.5cm}
\spav{1mm}\\
{\normalsize\em $^{(1)}$ Dipartimento di Scienze Fisiche,
Universit\`a di Napoli and INFN Sezione di Napoli,
Mostra d'Oltremare, Pad. 19, I-80125 Napoli, Italy.\\}
\spav{1mm}\\
{\normalsize\em $^{(2)}$ Centre de Physique Th\'eorique et de Mod\'elisation,
CNRS$^{~\dagger\dagger}$, Universit\'e de
Bordeaux I, \\ 19 rue du Solarium, F-33175 Gradignan,
France.\\}

\spav{5mm}

{\small\bf ABSTRACT\\}
\spav{2mm}\\
{\parbox{13cm}{\spao{4mm}
The BRST anomaly which may be present in the induced $W_n$ gravity quantized
on the light-cone is evaluated in the geometrical framework of Zucchini. The
cocycles linked by the cohomology of the BRST operator to the
 anomaly are straightforwardly calculated thanks to the analogy between this
formulation and the Yang-Mills theory. We give also a conformally covariant
formulation
 of these quantities including the anomaly, which is valid on arbitrary
Riemann surfaces. The example of the $W_3$ theory is discussed and a
comparison with other candidates for the anomaly available in the
literature is presented.

}\\}
\end{center}
\vfill

\noindent
{\footnotesize $^{\dagger}$ Partially supported by INFN and MURST 60\%.}

\noindent
{\footnotesize $^{\dagger\dagger}$ Unit\'e Associ\'ee au CNRS, U.R.A. 1537.}

\newpage

1.Introduction

In this letter we study the cohomology linked
 to the consistent anomaly possibly occuring in the induced $W_n$ gravities.
They are higher spin generalizations of 2-dim gravity whose symmetries
are the $W-$algebras, just as the Virasoro algebra appears as the residual
symmetry after gauging of 2-dim induced gravity.
 For recent reviews see \cite{r1,r2}.
 This anomaly and the cocycles linked to it by this cohomology are calculated
using Zucchini's formalism \cite{Z}. Recently, this author has
observed that
the generalized Beltrami differentials and projective connections which are
dynamical fields appearing in the induced light-cone $W_n$ gravity are
geometrical objects parametrizing in one-to-one fashion generalized
projective structures on a given Riemann surface $\Sigma$. This approach
allows us to realize the $W_n$ symmetry in terms of an explicit
parametrization of a 2-dim Yang-Mills (Y-M) connection. Thus the well-known
 cohomology of Y-M leads straightforwardly to the formulation of the
cohomology of any $W_n$-model. After a general presentation of this geometric
 approach, we give explicit results in the case of $W_3$ theory and compare
with a previously known candidate for the anomaly \cite{OSSV} in that case.
   \\

2. Zucchini's approach.

 We recall that an atlas of projective coordinates on a Riemann surface
$\Sigma$ is a collection of local homomorphisms $Z$ of $\Sigma$ into $C^1$
which are glued on overlapping domains by a Moebius transformation. Such an
atlas defines a complex structure on $\Sigma$, or equivalently conformal
classes of metrics which are related to the reference structure $z$ by the
Beltrami coefficient $\bc$ via Beltrami's equation. This projective structure
 $A$ is
in one-to-one correspondence with the pairs ($\bc$, $\ssc$) where $\ssc$ is
the projective connection (the Schwarzian derivative of $Z$). The algebra
which underlies this framework is the famous Virasoro algebra, corresponding
to the $W_2$ case.

 The approach \cite{Z} of the algebra $W_n$, where $n$ indicates the highest
spin of the generators involved, is based on a straightforward generalization
 of the notion of projective coordinates. It consists in enlarging the set of
coordinates by considering a collection of
local maps ($Z^1,...,Z^{n-1}$) of $\Sigma$ into $C^{n-1}$($Z^1 \equiv Z$)
\footnote{ We
consider only the holomorphic sector of the theory, all results derived
below being transposed to the antiholomorphic sector by complex
conjugation.}. On overlapping domains $K_{\alpha}$, $K_{\beta}$ the
transition functions of these local maps $Z_{\alpha}^i$ define a
$Sl(n;I\!\!\!C)$-valued 1-cocycle $\Phi_{\alpha \beta}$ on $\Sigma$ which in
turn defines a flat $Sl(n;I\!\!\!C)$ vector bundle $\Phi$ on $\Sigma$. Such
bundle is canonically associated to the generalized projective structure $A$.
These new data on the Riemann surface are used to build a basic object,
the matrix W, with entries

\begin{equation}
\label{1w}
     W_i^{\hspace{.05in}r}=\drp^i(\Delta^{-\frac{1}{n}}Z^r)
\end{equation}

where $\drp_z \equiv \drp$ and $\Delta=$det$ \parallel \drp^p Z^l
\parallel $. One can easily verify that det$W=1$.
Under a holomorphic coordinate change on $\Sigma$ and a map change of the
projective atlas, one has

\begin{equation}
\label{w00}
W_{b\beta}=\Lambda_{ba}W_{a\alpha}(\Phi^{-1})^t_{\alpha \beta}
\end{equation}

where $\Phi_{\alpha \beta}$ is the $Sl(n;I\!\!\!C)$ cocycle defined above and
 $\Lambda_{ab}$ are the transition matrices of the bundle of ($n-1$)-jets of
sections of $k^{(1-n)/2}$, where $k$ is the canonical line bundle
with transition functions $k_{ba} \equiv \drp_b z_a$.

One further define the two matrices

\begin{equation}
\label{w0}
\om=\drp W W^{-1} \hspace{.5in} ; \hspace{.5in} \oms=\drpb W W^{-1},
\end{equation}

which are traceless and independent of the choice of a chart in the
projective atlas. They glue under conformal coordinate changes as follows:

\begin{equation}
\label{g0}
\om_b=k_{ba}(\Lb\om_a\Lb^{-1}+\drp\Lb\Lb^{-1}),
\end{equation}

\begin{equation}
\label{g1}
\oms_b=\bar{k}_{ba}(\Lb\oms_a\Lb^{-1}).
\end{equation}

The Wronskian form of $W$ implies that the elements of $\om$ are trivial
(zero or one) except $n-1$ elements $\rho_i \equiv \om_{n-1}^i$ which
characterize the maps
($Z^1,...,Z^{n-1}$). By taking suitable combinations of these objects and
their derivatives \cite{SG,zuber} new fields ($\tilde{\rho}_i$; $i=1,..,n-2$;
 $i\neq 0$) are constructed which have definite conformal spins $i+2$.
 Moreover
the field $\tilde{\rho}_{n-2}$ is a projective connection transforming as

\begin{equation}
\tilde{\rho}_{{n-2}_b}=k^2(\tilde{\rho}_{{n-2}_a}+
\frac{n(n^2-1)}{12}S(z_b,z_a))
\end{equation}

where $S(z_b,z_a)$ is the Schwarzian derivative given by
$$ S(z_b,z_a)=\drp_a^2 ln\drp_a z_b-\frac{1}{2}(\drp_a ln\drp_a z_b)^2.$$
 The matrices $\om$ and $\oms$ can be considered as the two components of a
2-dim flat connection since they satisfy the relation

\begin{equation}
\label{g2}
\drpb\om-\drp\oms+[\om,\oms]=0.
\end{equation}

These flatness conditions first allow us to determine elements of $\oms$ in
terms of $\om$ and of $n-1$ fundamental fields
$\mu_i \equiv {\oms}_i^{n-1}$ ($i=1,..,n-1$) which
 are the generalizations of the usual Beltrami coefficient. They also give
the holomorphic conditions obeyed by the $\tilde{\rho}_i$. For $n=2$ these
conditions reduce to the unique relation

\begin{equation}
(\drpb-\mu_0\drp -2\drp \mu_0)\rho_0=-\frac{1}{2} \drp^3 \mu_0,
\end{equation}

which is exactly the anomalous Ward identity of the induced chiral 2-dim
gravity, once the field $\rho_0$ has been identified with the induced
energy-momentum tensor and $\mu_0$ (the external source coupled to it) with
the Beltrami differential on $\Sigma$.

The relation between 2-dim induced gravity and a 2-dim partially gauge fixed
$Sl(2;I\!\!R)$ Y-M theory was first found by Polyakov \cite{Pol}. In ref.
\cite{Y},
 the generalization of Polyakov's gauge-fixing on an arbitrary Riemann
surface (see also ref.\cite{ZZ}) led to the introduction of $W$ matrices of
$Sl(2;I\!\!\!C)$ depending on a single projective coordinate. The
construction made
in \cite{Z} suggests the interpretation of the holomorphic conditions deduced
from (\ref{g2}) as the Ward identities of the induced $W_n$ gravity, once the
 correct identifications for the fields $\tilde{\rho}_i$ (and the
corresponding generalized Beltrami differentials $\tilde{\mu}_i$) have been
made.
   The association of a zero curvature condition to the formulation of
induced $W_n$ gravity and its interpretation as an anomaly equation are not
new \cite{Bilal,Das}. The essential advantage of this geometrical
framework is to define the $W_n$ symmetries as gauge
transformations of the vector bundle $\Phi$  and to provide a systematic
method to derive the corresponding nilpotent BRST algebra, as we now discuss.
\\

3.BRST symmetry

The vector bundle $\Phi$ can be considered as a functional of the fields
($\tilde{\rho}_i,\tilde{\mu}_i$); the variations of these fields leaving this
 bundle invariant are precisely the form of the $W_n$ symmetry
transformations. They are obtained from infinitesimal variations of the maps
$Z$ of $A$ \cite{Z}

\begin{equation}
\delta Z^r=\epsilon^r_s Z^s-\epsilon^0_s Z^rZ^s
\label{it}
\end{equation}

 which when written in terms of a ghost matrix field $\gamma$, instead of
infinitesimal parameters $\epsilon$, give the corresponding BRST
transformations. The matrix $\gamma$ has zero trace and transforms under
changes of trivialization of the bundle $\Phi$ as

\begin{equation}
\gamma_{\beta} \Phi_{\beta \alpha}=\Phi_{\beta \alpha} \gamma_{\alpha}.
\label{gt}
\end{equation}

Geometrically, this means that the matrix function $\gamma$ corresponds to a
finite gauge transformation of $\Phi$. This allows to construct \cite{Z} with
 the help of $\gamma$ and $W$ a traceless matrix $C$ (which has a ghost
grading one)
which is independent of map choices in the projective structure.
 Moreover, under a conformal coordinate change in the canonical
bundle $k$ this ghost matrix transforms as

\begin{equation}
C_{b}= \Lambda_{ba}C_{a} \Lambda_{ba}^{-1}.
\label{w1a}
\end{equation}

Now nilpotency of the law (\ref{it}) is fulfilled if $s\gamma = \gamma^2$
and the BRST transformation of the matrix $W$ is then

\begin{equation}
sW=CW.
\label{w1}
\end{equation}

By considering this equation as the definition of $C$, it appears that
$C$ is thus not only a function of the ghosts $\gamma$ but also
of the variables $Z^i$ (in fact the $\rho_i$). Its explicit form can be
deduced from (\ref{1w}) and will not be given here. It can be found in
ref.\cite{Z}.

Moreover

\begin{equation}
sC= CC.
\label{w2}
\end{equation}

Obviously this transformation is nilpotent and looks like the well-known BRST
 transformation of the Faddeev-Popov ghost in Y-M theory.
{}From eq.(\ref{w1}) and the definition (\ref{w0}) of the matrix $\om$ one has

\begin{equation}
s\om=\drp C+[C,\om].
\label{w3}
\end{equation}

Analogously it is straightforward to deduce the BRST transformation of $\oms$

\begin{equation}
s\oms=\drpb C+[C,\oms].
\label{w4}
\end{equation}

The parallel with the Y-M formalism leads us to gather together the two
matrices $\om$ and $\oms$ into a 2-dim gauge connection

\begin{equation}
{\cal A} = \om dz+\oms d\zb,
\label{w5}
\end{equation}

which transforms as ($d$ being the usual external derivative)

\begin{equation}
s{\cal A} = -dC+[{\cal A},C].
\label{w6}
\end{equation}

 The curvature (field strength) ${\cal F}=d\ca-\ca \ca$, corresponding to the
 connection $\ca$ is in fact zero due to the holomorphy condition (\ref{g2}).

It is well known that the Ward identities for induced $W$ gravity are very
similar in structure to the BRST transformations of the projective
connection and the currents. For 2-dim gravity (i.e $W_2$) this is
a direct consequence of the analogy between the
Beltrami equation $\drpb Z=\bc \drp Z$ and the BRST transformation
$sZ=c\drp Z$, where $c$ is the ghost introduced by Becchi \cite{Bec}.
Accordingly every function of the projective coordinate $Z$ and of
($\drp,z$) will exhibit this relation between its BRST transformation
and the relation induced by the Beltrami equation.
 The same sort of relation is
present in the $W_n$ models and was previously discussed in ref \cite{Boer}.
 This is obvious if we compare the transformation law
(\ref{w1}) with the definition (\ref{w0}) of the matrix $\os$ considered as
the holomorphic condition verified by $W$.
 In fact the relation (\ref{w3}) can be deduced
from the compatibility condition (\ref{g2}) between $\om$ and $\oms$ by
replacing the matrix $\oms$ and the partial derivative $\drpb$ by the ghost
matrix $C$ and the BRST operator $s$ respectively.

Thus the matrix $C$ is straightforwardly deduced from the matrix $\oms$ by
replacing the generalized Beltrami coefficients $\mu_i$ by the ghost fields
$c_i$ and consequently there are $n-1$ independent ghost fields
($c_i$; $i=1,..,n-1$) in $C$.
   \\

4.Consistent anomalies.

Now that the contact with the 2-dim Y-M theory has been made, the
calculation of the consistent anomaly is easy. The Chern-Simons action,

\begin{equation}
S=\int Tr({\cal A}d{\cal A}+\frac{2}{3}{\cal A}{\cal A}{\cal A}),
\end{equation}

which is defined on a three dimensional compact manifold, having $\Sigma$ as
boundary, can be viewed as the integral of a Chern-Weil characteristic
polynomial $T_3^0$ of rank 3. The lower index denotes the form degree and the
 upper index the ghost number. Consequently, by applying the BRST operator
(\ref{w6}) to this polynomial, we obtain the descent equations

\begin{eqnarray}
sT_3^0+dT_2^1&=&0,      \nonumber   \\
sT_2^1+dT_1^2&=&0,      \label{d1}  \\
sT_1^2+dT_0^3&=&0,      \label{d2}  \\
sT_0^3=0,             \label{d3}
\end{eqnarray}

where the explicit expressions of the cocycles are given by

\begin{eqnarray}
T_2^1&=&3Tr(C\om \os -C\os \om)dz d\zb,     \label{f1}  \\
T_1^2&=&3Tr(C^2\om dz+C^2\os d\zb),      \label{f2}  \\
T_0^3&=&Tr(C^3).      \label{f3}
\end{eqnarray}

It is obvious that a cohomological analysis can determine the anomaly only
modulo a (possibly vanishing) multiplicative constant whose value can be
calculated, for a given model, through other methods, such as Feynman
diagrams computation. Unless otherwise stated, what we call anomalies in this
 cohomological analysis have to be understood as possible candidates.
They are non-trivial solutions of the Wess-Zumino consistency condition
\cite{WZ}.

In fact the expressions above are valid only on the plane and on the sphere.
Generalization to a Riemann surface of higher genus has been made only in the
 case of the bosonic string \cite{LS} and in the $W_3$ case \cite{Boer} by
introducing by hand a projective connection which is holomorphic and BRST
inert. The present formalism allows us also to achieve a formulation of the
anomaly associated to a given $W_n$-model which is well-defined on $\Sigma$.
 This formulation has the advantage of being self-contained, since the fields
used to render the local expressions valid on any local coordinate chart on
$\Sigma$ are the field $\tilde{\rho}_{n-2}$ which, under a
conformal change of coordinates transforms as a projective connection,
and the projective invariants which transform
homogeneously, i.e are sections of $k^{n-i}$.

Instead of calculating the anomaly and its cocycles by using the Chern-Weil
characteristic polynomial, it is also possible to start from the well-known
2-dim Y-M (non-integrated) anomaly $Tr(Cd{\cal A})$, resulting in

\begin{equation}
\label{10a}
a^1_2=3Tr[C(\drp \os-\drpb \om)]dzd\zb.
\end{equation}

This expression of the anomaly is shown to be equivalent to (\ref{f1})
thanks to the pure gauge condition (\ref{g2}). Moreover one can show that
the 2nd term in the r.h.s of (\ref{10a})
is a well-defined density on $\Sigma$ so that the integration makes sense.

 For later convenience let us consider (in fact as we have verified on the
explicit examples $W_2$ and $W_3$ this choice is the most
economical one)

\begin{equation}
\label{11}
\sigma^1_2=3Tr(\oms s\om-C\drpb \om)dzd\zb.
\end{equation}

Indeed, using the transformation laws (\ref{g0},\ref{g1},\ref{w1a}) and
noting that by construction $\drpb \Lb=0$, we can easily verify that
$\sigma^1_2$ is a well-defined 2-form. This expression  of the anomaly is
related to the formula (\ref{f1}) given by the Y-M formalism by

\begin{equation}
\label{12}
\sigma^1_2=T^1_2-3Tr(\drpb (C\om)-s(\oms \om))dzd\zb.
\end{equation}

Then a well-defined system of descent equations is given by the following
 cocycles

\begin{eqnarray}
\sigma_2^1&=&3Tr(\os\drp C-C\drp\os+2C(\om\os-\os\om))dzd\zb,  \label{13a}  \\
\sigma_1^2&=&-3Tr(C\drp C+2C^2\om)dz +Tr(\os C^2)d\zb,     \label{13}  \\
\sigma_0^3&=&Tr(C^3),      \label{14}
\end{eqnarray}

where the expression (\ref{13a}) is obtained by substituting in (\ref{11})
$s\om$ and $\drpb\om$ by (\ref{w3}) and (\ref{g2}) respectively.
 Thus the formalism presented above provides us
with a completely algorithmic procedure for calculating the consistent and
conformally covariant anomalies associated to a given $W_n$-model and the
cocycles related to these anomalies by the system of descent equations.

The covariant version of the differential operators $\drp^n$, the so-called
Bol operators
$L_n$ are well known \cite{gierop}. Since the BRST transformations of the
fields are also covariant quantities, these operators will appear naturally
both in the
formulation of these laws and in the expression of the covariant cocycles and
 anomaly. For later reference we display the explicit expressions of $L_3$
and $L_5$.

\begin{eqnarray}
L_3(R)&=&\drp^3+2R \drp +(\drp R)     \label{B1}  \\
L_5(R)&=&\drp^5+10R \drp^3 +15( \drp R) \drp^2 +[9(\drp^2 R)+16R^2]\drp
+2[(\drp^3 R)+8R(\drp R)]         \label{B2}
\end{eqnarray}

 Let us stress that the Bol operator $L_n$ is only covariant when acting on a
 conformal field of conformal dimension ($\frac{(1-n)}{2}$) in the
holomorphic sector, provided $R$ transforms with the Schwarzian derivative
under a conformal change of coordinates:

\begin{equation}
R_b=k^2(R_a-S(z_b,z_a)).
\end{equation}
  \\
5. The induced gravity $W_3$.

We illustrate this construction for the $W_3$ algebra. First a technical
problem appears for $n \geq 3$: the geometrical fields parametrizing the
matrix $W$ are not the physical fields since in general they do not transform
 homogeneously when one goes from one chart to another. Combinations $H$ of
these fields which are sections of the fiber bundle
$k^p, \! \bar{k}k^{-q}$ have to be considered; they transform as differentials
 under a conformal change of coordinates

\begin{equation}
H_b=H_a k^h \bar{k},
\label{d10}
\end{equation}

where $h=p$ or $-q$ is the conformal weight of the field $H$ \footnote{In
sect.2 these fields were denoted by a tilde superscript which is omitted here
to simplify the notations.}.

The sets of pairs of such fields used in the $n=3$ case are the generalized
Beltrami coefficients ($\bc,\bbc$), the projective connection $\ssc$ and the
associated projective invariant $\rz$ and the ghosts ($\cz,\czz$).
Now the components of the connection ${\cal A}$ are

\begin{equation}
\om=\left(\begin{array}{ccc}
             0 & 1 & 0 \\
         0  & 0 & 1 \\
         \rz & \ssc & 0

      \end{array}
      \right)
\label{b3}
\end{equation}

\begin{equation}
\oms=\left(\begin{array}{ccc}
             \ds \drp^2 \bbc-\f{2} \bbc \ssc-\drp \bc & \bc- \dd \drp \bbc
& \bbc  \\
   &  &   \\
           \drp \oms_{00}-\dd \bbc \drp \ssc   & -\dt (\drp^2 \bbc-\bbc \ssc)
 &  \bc+ \dd \drp \bbc   \\
   &   &  \\
 \drp\oms_{01}+\drp(\bbc \rz)+  & \dd\drp\oms_{11}
-\drp^2\bc+\bc\ssc  & \drp\oms_{12}+\oms_{11}  \\
\bc\rz+\dd \bc\drp \ssc  &
+\bbc\rz  &
     \end{array}
      \right),
\label{b44}
\end{equation}

whereas the ghost matrix is deduced from the $\os$ matrix by substituting for
the generalized Beltrami coefficients $\bc$, $\bbc$, the ghosts $\cz$ and
$\czz$ respectively.

\begin{equation}
C=\left(\begin{array}{ccc}
             \ds \drp^2 \czz-\f{2} \czz \ssc-\drp \cz & \cz- \dd \drp \czz
& \czz  \\
   &  &  \\
           \drp C_{00}-\dd \czz \drp \ssc   & -\dt (\drp^2 \czz-\czz \ssc)
 &  \cz+ \dd \drp \czz   \\
   &  &  \\
 \drp C_{01}+\drp(\czz \rz)+  & \dd\drp C_{11}
-\drp^2\cz+\cz\ssc  & \drp C_{12}+C_{11}  \\
 +\cz\rz+\dd \cz\drp \ssc  &+\czz\rz  &
     \end{array}
      \right).
\label{b5}
\end{equation}

Projecting down the matrix equations (\ref{w2},\ref{w3}) to the components
(0,1) and (0,2) for instance, and (\ref{w4}) to (2,0) and (2,1) we obtain
the automatically nilpotent laws

\begin{eqnarray}
s\czz&=&2\czz \drp \cz- \drp \czz \cz,   \label{s31} \\
s\cz&=&\cz \drp \cz-\ds \czz \drp^3 \czz+\frac{1}{4}\drp\czz\drp^2\czz-
\f{2} \drp\czz\czz\ssc,   \label{s32}    \\
s\bbc&=&\cd_{-2}(\bc)\czz+\cz \drp \bbc-2\bbc \drp \cz,  \label{s33}  \\
s\bc&=&\cd_{-1}(\bc)\cz-\frac{1}{12}(2\czz \drp^3-3\drp \czz \drp^2 +
3\drp^2\czz \drp-2\drp^3\czz )\bbc  \nonumber  \\
&+&\f{2}(\czz \drp \bbc-\bbc \drp \czz)\ssc,
  \label{s23}  \\
s\ssc&=-2&L_3(-\dd \ssc) \cz+(2 \czz \drp \rz+3 \drp \czz \rz),
\label{s24} \\
s\rz&=&\frac{1}{6}L_5(-\dd \ssc) \czz+3\rz \drp \cz+\cz \drp \rz,
\label{s25}
\end{eqnarray}

where we have introduced the following derivative acting on any field of
conformal dimension $J$ as:

\begin{equation}
\cd_J(\bc)=\drpb -\bc \drp -J \drp \bc.
\label{der}
\end{equation}

These laws have to be complemented by the constraints  of
the holomorphic conditions
\begin{eqnarray}
\bar \partial\ssc&=- 2&L_3(-\dd \ssc) \bc+(2 \bbc \drp \rz+3 \drp \bbc \rz),
\label{s24holom} \\
\bar\partial \rz&=&\frac{1}{6}L_5(-\dd \ssc) \bbc+3\rz \drp \bc+\bc \drp \rz,
\label{s25holom}
\end{eqnarray}
which express, in the parametrization (\ref{b3}, \ref{b44}) of $\Omega$ and
$\Omega^*$, the flatness of the connection. One should notice that the BRST
transformation of the $\rho$ fields can be obtained from (\ref{s24holom},
\ref{s25holom}) through the replacement of $\bar \partial$ by the operator
$s$ and of $(\bc,\, \bbc)$ by $(\cz,\, \czz)$, as discussed at the end of
Sect.3.

The covariant non-integrated $W_3$ anomaly is now obtained by replacing
(\ref{b3},\ref{b44},\ref{b5}) in (\ref{13a}). Modulo  total
derivatives, it reads

\begin{eqnarray}
\sigma_2^1&=&[\ds(\bbc L_5(-\dd \ssc) \czz-\czz L_5(-\dd \ssc) \bbc)-
2(\bc L_3(-\dd \ssc) \cz  \nonumber   \\
&-&\cz L_3(-\dd \ssc) \bc) +2\rz (\drp \czz \bc- 2\drp \bc\czz+2\drp \cz
\bbc- \cz \drp\bbc)]dzd\zb.
\label{71}
\end{eqnarray}

Using (\ref{s24holom},\, \ref{s25holom}) the covariant anomaly
(\ref{71}) takes the compact form (compare
with (\ref{11})) $$\sigma=\mu s\rho- c\drpb \rho$$ thus proving the overall
consistency of our results \footnote{A similar expression for $\sigma_2^1$
has been obtained in ref.\cite{GGL}.}.

A different form of the $W_3$ anomaly has been found by the authors of
\cite{OSSV}, as solution of the Wess--Zumino consistency conditions. It is
easy to show that the BRST operator corresponding to their transformation
algebra coincides with that given by eqs.(\ref{s31}--\ref{s25}). Their
expression for the anomaly, rewritten in terms of our fields reads

\begin{equation}
\displaystyle{\tilde \sigma_2^1=
\,  \{-{1\over 3}\,{\it c^{zz}}\, \left(\, \partial^5
\mu^{zz}_{\bar z} -
 4\, (\, \rho_{zz}^2 \, \partial \mu^{zz}_{\bar z} +
\rho_{zz} \, \partial \rho_{zz} \, \mu^{zz}_{\bar z} \,)\, \right) +
4\, {\it c^z} \, \partial^3 \mu^z_{\bar z}  \} dz d\bar z
}
\label{AOoguri}
\end{equation}
We have explicitly checked that both expressions (\ref{71}) and
(\ref{AOoguri}) are in fact cocycles of the BRST algebra ( once the
holomorphy conditions are taken into account). Accordingly they have to be
equivalent, and, in fact, the following relation holds (again modulo the
holomorphy conditions)

\begin{equation}
\sigma^1_{z \bar z}+\tilde \sigma^1_{z \bar z}=\bar\partial\,
{\it \Delta}_{z}^1+
\partial\,{\it \Delta}_{\bar z}^1 - {\it s}\,{\it \Delta}_{z \bar z}^0.
\label{aequiv2}
\end{equation}

Since this equation is invariant under the transformations
\begin{equation}
{\it \Delta}_{z\bar z}^0\rightarrow {\it \Delta}_{z\bar z}^0+\partial
 u_{\bar z} -\bar\partial u_z;\quad
{\it \Delta}_{z}^1\rightarrow {\it \Delta}_{z}^1+\partial v -s\, u_z;\quad
{\it \Delta}_{\bar z}^1\rightarrow {\it \Delta}_{\bar z}^1+\bar\partial v
-s\, u_{\bar z};
\end{equation}
the general form of the $\it \Delta$'s is given by
\begin{equation}
\begin{array}{l}
\displaystyle{
{\it \Delta}_{z \bar z}^0 =
- 2\,{\mu^{zz}_{\bar z}} \, {\rho_{zzz}}
- 2\,{\mu^z_{\bar z}} \, {\rho_{zz}} +
\partial  u_{\bar z} - \bar\partial u_z
}\\
    \\
\displaystyle{
{\it \Delta}_{z}^1 =
-2\,{\it c^{zz}} \, {\rho_{zzz}}
-2\,{\it c^z} \, {\rho_{zz}} +
\partial v - s\, u_z
}
\\
   \\
\displaystyle{
{\it \Delta}_{\bar z}^1 =
4\,(c^z\,\partial^2\mu_{\bar z}^z-\partial c^z\,\partial \mu_{\bar z}^z+
\partial^2 c^z\,\mu_{\bar z}^z)-
}\\
    \\
\displaystyle{
\frac{1}{3}(\partial^4 c^{zz}\,\mu_{\bar z}^{zz}-
\partial^3 c^{zz}\,\partial\mu_{\bar z}^{zz}+
\partial^2 c^{zz}\,\partial^2\mu_{\bar z}^{zz}-
\partial c^{zz}\,\partial^3\mu_{\bar z}^{zz}+
 c^{zz}\,\partial^4\mu_{\bar z}^{zz})+
}\\
    \\
\displaystyle{
\frac{1}{3}(
4\,\partial^2 c^{zz}\,\mu_{\bar z}^{zz}-
\frac{5}{2}\partial c^{zz}\,\partial\mu_{\bar z}^{zz}+
c^{zz}\,\partial^2\mu_{\bar z}^{zz})\,\rho_{zz}-
\frac{1}{3}(
 c^{zz}\,\partial\mu_{\bar z}^{zz}-
\frac{7}{2}\partial c^{zz}\,\mu_{\bar z}^{zz})\,\partial\rho_{zz}+
}\\
    \\
\displaystyle{
\frac{1}{3}\, c^{zz}\,\mu_{\bar z}^{zz}\,\partial^2\rho_{zz} -
2c^z\mu_{\bar z}^z\,\rho_{zz} + 2\,(c^{z}\,\mu_{\bar z}^{zz} -
 c^{zz}\,\mu_{\bar z}^{z})\rho_{zzz}
+ \bar\partial v - s\, u_{\bar z}}
\end{array}
\label{***}
\end{equation}
where $u_z$, $u_{\bar z}$, are the holomorphic,
antiholomorphic components of an
arbitrary $1$-form with ghost number $0$ ( the inspection of possible terms
fixes $u_z=0$ and $u_{\bar z}$ to be an
arbitrary linear combination of the monomials
$\partial \mu_{\bar z}^z$, $\partial^2 \mu_{\bar z}^{zz}$ and $\mu_{\bar
z}^{zz}\rho_{zz}$) and $v$ is an an arbitrary $0$-form with ghost number
$1$ (so an arbitrary linear combination of
$\partial c^z$, $\partial^2 c^{zz}$ and
$ c^{zz}\rho_{zz}$). Evidently, the relation between the (integrated)
anomalies is independent of the choice of $u_{\bar z}$ and $v$ since we can
disregard ${\it \Delta}_z^1$, ${\it \Delta}_{\bar z}^1$ and the term
involving $u_{\bar z}$ in $\Delta_{z \bar z}^0$ because arising from total
derivatives.

It should be noticed that both expressions (\ref{71}) and (\ref{AOoguri})
contain, besides the leading terms $\mu_{\bar z}^z\partial^3 c^z$ and
$\mu_{\bar z}^{zz}\partial^5 c^{zz}$, additional contributions which are
necessary to insure the consistency of the anomaly.

Leading terms of the form $\mu_{\bar z}^{z\dots z}\partial^{2n-1} c^{z\dots
z} $ (called universal anomalies by Hull \cite{Hull, r2}) are present in
all models obtained by coupling a system of free scalar fields to $W_n$
gravity\footnote{
The anomalous Ward identities in the $W_n$ case have been already
discussed in refs.\cite{Bilal,Das}.}
 \cite{Hull,SSN}. Evidently, conformal invariance requires the
occurrence of additional terms that together with Hull's universal terms
lead to the necessary Bol operators. Thus, these terms are
easily obtained. On the other side, as discussed above, additional terms,
 involving the higher spin
$\rho$ fields, are also needed, which cannot be easily guessed. However, in
our approach the general expressions for $C$, $\Omega$ and $\Omega^*$ given
in \cite{Z} and eq. (\ref{11}) allow us to obtain straightforwardly
them by a mere repetition of the computations.
    \\

     We would like to thank F.Biet, Y.Noirot  and R.Zucchini for valuable
discussions.
     \\

Note added: after submition of the present paper, it appeared a work where is
 discussed a comparison between the anomalies of ref.\cite{GGL} and
ref.\cite{OSSV}, see P. Watts, Generalized Wess--Zumino consistency conditions
 for pure ${\cal W}_3$ anomalies. CPT-95/P.3237; hep-th/9509044.

\end{document}